\documentclass{article}
\usepackage{spconf,amsmath,graphicx}
\usepackage{microtype}
\usepackage{doi}
\usepackage{booktabs}
\usepackage{float} 
\usepackage{mathrsfs}
\usepackage{subfigure}
\usepackage{threeparttable}
\usepackage{multicol}
\usepackage{multirow}
\usepackage{algorithmic}
\usepackage{textcomp}
\usepackage{xcolor}
\usepackage{amssymb}
\usepackage{url}
\usepackage{pifont}
\usepackage{enumitem}
\setlist{nosep, leftmargin=14pt}

\usepackage{mwe} % to get dummy images

\title{Reconstruction of knee MRI from one corresponding X-ray via deep learning}
\name{Zhe Wang$^{1}$ \qquad Aladine Chetouani$^{2}$  \qquad Rachid Jennane$^{1\star}$}
\address{$^{1}$ IDP Institute, UMR CNRS 7013, University of Orleans, France\\ $^{2}$PRISME Laboratory, EA 4229, University of Orleans, France}

\begin{document}
%\ninept
%
\maketitle
\begin{abstract}
Generally, X-ray, as an inexpensive and popular medical imaging technique, is widely chosen by medical practitioners. With the development of medical technology, Magnetic Resonance Imaging (MRI), an advanced medical imaging technique, has already become a supplementary diagnostic option for the diagnosis of KOA. We propose in this paper a deep-learning-based approach for generating MRI from one corresponding X-ray. Our method uses the hidden variables of a Convolutional Auto-Encoder (CAE) model, trained for reconstructing X-ray image, as inputs of a generator model to provide 3D MRI.
\end{abstract}
\begin{keywords}
Convolutional Auto-encoder, Knee osteoarthritis, MRI, X-ray
\end{keywords}
\section{Introduction}
\label{sec:intro}
Deep learning has gained traction in the field of computer vision in recent years, particularly in the usage of Convolutional Neural Networks (CNNs). In \cite{zhe}, Wang et al. used two patches from the lateral and medial parts of the knee joint X-ray images as inputs of the proposed Siamese-GAP network. Meanwhile, 3D CNN, which is developing rapidly and gaining interest as a method for analysing image sequences, has been shown to be useful in medical image processing. In \cite{lung}, the use of 3D volumetric data in a 3D CNN outperformed 2D CNN when classifying lung nodules. In this paper, we propose a deep-learning-based approach for generating the MRI sequence images from one single X-ray. More specifically, through the hidden variables between the encoder and decoder, the generator module learns the discriminative features embedded in a single 2D X-ray and transforms them into 3D MRI sequence images.

\begin{figure*}
\centering  %图片全局居中
\includegraphics[width=0.8\textwidth]{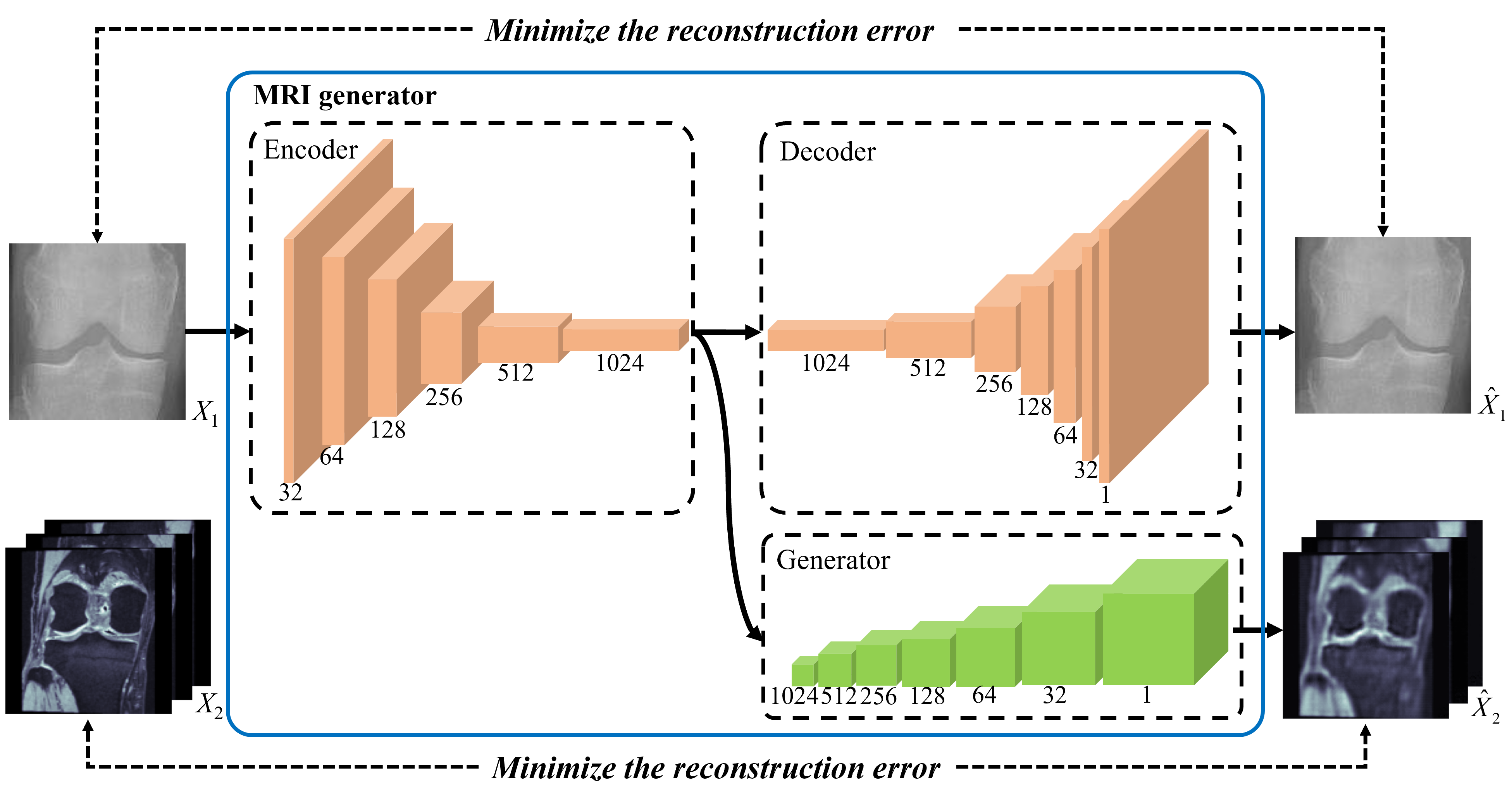}
\caption{The architecture of our approach, which consists of three parts: an encoder, a decoder, and a generator. Orange blocks in the encoder and decoder represent 2D CNN and 2D DCNN, respectively. Green blocks in the generator represent 3D DCNN.}
\label{fig-flowchart}
\end{figure*}

\section{Proposed Approach}

We propose in this study to exploit the features given by an AE model for X-ray images to generate the corresponding MRI. As illustrated Fig. \ref{fig-flowchart}, the overall architecture of our method is composed of three main modules: an encoder, a decoder, and a generator. Each of these modules is presented in this section.

\textbf{Encoder:} It aims to extract relevant features from an X-ray image. It consists of six 2D CNNs blocks of a 2D Convolutional layer of the same kernel size of 3$\times$3 with different depths (i.e., 32, 64, 128, 256, 512, and 1024) - a Batch Normalization (BN) layer - a Leaky Rectified Linear Unit (LeakyReLU) layer with a predefined slope of 0.2.

\textbf{Decoder:} It aims to reconstruct the X-ray image through the extracted features (i.e. encoder) and forces the model to give a relevant feature representation of the X-ray image. It has an almost symmetrical structure to the encoder. It consists of seven 2D DeCovolutional Neuronal Network (2D DCNN) blocks of a 2D Deconvolutional layer of the same kernel size of 4$\times$4 with different depths (i.e., 1024, 512, 256, 128, 64, 32, and 1) - a BN layer - a LeakyReLU layer with a predefined slope of 0.2.

\textbf{Generator:} It consists of seven 3D DCNN and two 2D CNN blocks. Each 3D DCNN block is composed of a set of a 3D deconvolutional layer of the same kernel size of 4$\times$4$\times$4 with different depths (i.e., 1024, 512, 256, 128, 64, 32, and 1) - a 3D BN layer - a ReLU layer. The first 2D CNN block is  composed of a 2D convolutional layer of the kernel size of 1$\times$1 with the depths of 8 - a BN layer - a ReLU layer, while the second one is made up of a 2D convolutional layer of the kernel size of 1$\times$1 with depths of 128 with an output channel of 20.

\subsection{Hybrid loss strategy}

To ensure that the model can extract the necessary features for X-ray image reconstruction and generate MRI images efficiently, the loss function used in this study consider both outputs (i.e. $\hat{X_1}$ and  $\hat{X_2}$). To do so, we propose the following hybrid MSE loss function $J_{hybrid}$ :

\begin{equation}
\label{MSE}
J_{hybrid} = \lambda_1 J_{MSE}(X_1, \hat{X_1}) + \lambda_2 J_{MSE}(X_2, \hat{X_2})
\end{equation}
where the hyper-parameters $\lambda_1$ and $\lambda_2$ are used to weigh two reconstruction losses.

\section{Experimental settings}

\subsection{Knee database}
In this study, a freely-accessible database, OsteoArthritis Initiative (OAI) \cite{OAI}, was used. The OAI is a longitudinal research that includes nine follow-up assessments of 4,796 participants over 96 months. Participants range in age from 45 to 79 years. The OAI seeks to enrol participants with KOA or a high risk of developing it. 

\subsection{Data preprocessing}
\label{data_preprocessing}
\begin{figure}[htbp]
\centering
\subfigure[]{
\label{original}
\begin{minipage}[t]{0.13\textwidth}
\centering
\includegraphics[width=1\textwidth]{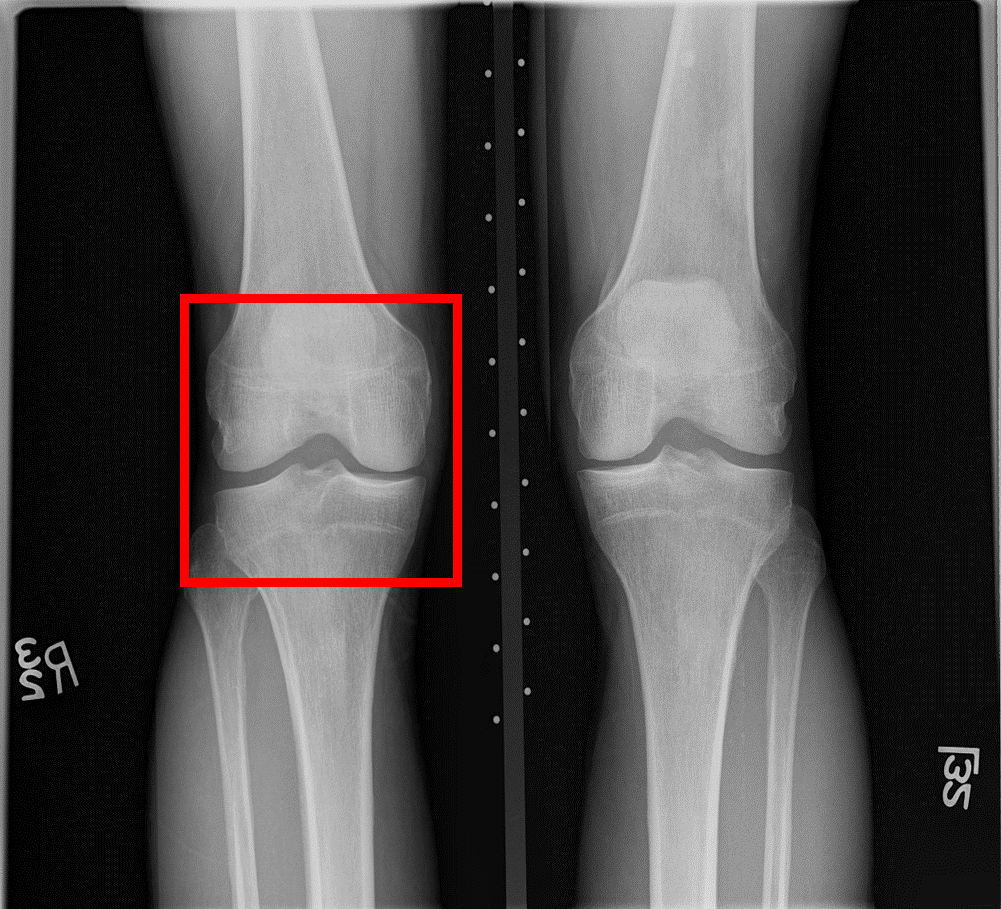}
%\caption{a typical knee X-ray}
\end{minipage}
}
\subfigure[]{
\label{ROI_Xray}
\begin{minipage}[t]{0.12\textwidth}
\centering
\includegraphics[width=1\textwidth]{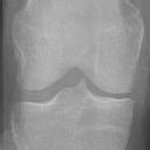}
\end{minipage}
}
\subfigure[]{
\label{ROI_MRI}
\begin{minipage}[t]{0.16\textwidth}
\centering
\includegraphics[width=0.8\textwidth]{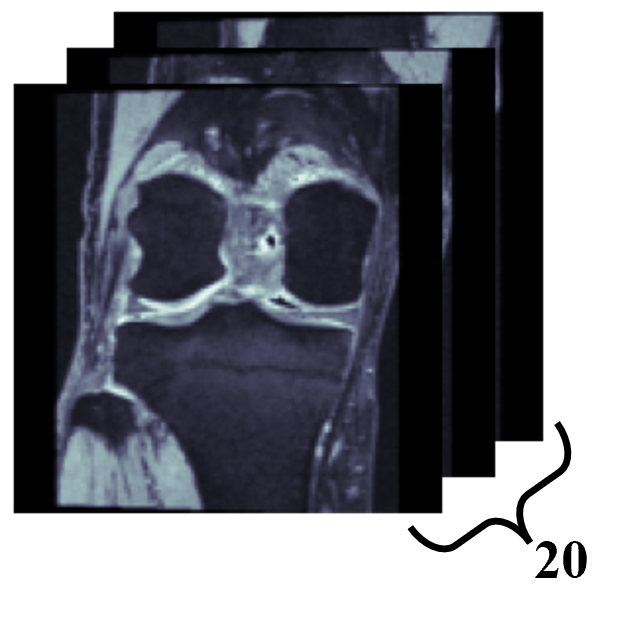}
\end{minipage}
}
\caption{(a) Original knee X-ray, (b) obtained ROI in red, and (c) obtained MRI series.}
\label{Fig-Patches}
\end{figure}

The ROIs extracted using YOLOv2 (see Fig. \ref{original} and \ref{ROI_Xray}) were employed as inputs of the encoder module. For the MRI series, the average number of slices for each corresponding MRI is 65. The most useful slices are always between the 20th and the 40th after the visual evaluation. Therefore, for each X-ray input, its corresponding MRI has 20 slices. It is noteworthy that the inputs (i.e., X-ray and each slice of MRI) are resized to 128 $\times$ 128 pixels.

\subsection{Experimental details}
The Adam optimizer was used for training with 1,000 epochs, a mini-batch of size 32, and learning rates of 0.001 and 0.0001 for the encoder-decoder and generator, respectively.

\section{Acknowledgment}
This work was supported by the French National Agency of Research (ANR) through the ANR-20-CE45-0013-01 project.

\bibliographystyle{IEEEbib}
\bibliography{strings,refs}

\begin{thebibliography}{1}

\bibitem{zhe}
Zhe Wang, Aladine Chetouani, Didier Hans, Eric Lespessailles, and Rachid
  Jennane,
\newblock ``Siamese-gap network for early detection of knee osteoarthritis,''
\newblock in {\em 2022 IEEE 19th International Symposium on Biomedical Imaging
  (ISBI)}, 2022, pp. 1--4.

\bibitem{lung}
Raunak Dey, Zhongjie Lu, and Yi~Hong,
\newblock ``Diagnostic classification of lung nodules using 3d neural
  networks,''
\newblock in {\em 2018 IEEE 15th international symposium on biomedical imaging
  (ISBI 2018)}. IEEE, 2018, pp. 774--778.

\bibitem{OAI}
G.~Lester,
\newblock ``{The Osteoarthritis Initiative: A NIH Public–Private
  Partnership},''
\newblock {\em HSS Journal: The Musculoskeletal Journal of Hospital for Special
  Surgery}, vol. 8, no. 1, pp. 62--63, 2011.

\end{thebibliography}

\end{document}